\documentclass[aps,pre,twocolumn,superscriptaddress,showkeys,showpacs,amsfonts,amssymb,amsmath]{revtex4}

\usepackage{bm}
\usepackage{graphicx}
\usepackage{subfigure} 

\begin{document}

\newcommand{\fig}[1]{Fig.~\ref{fig:#1}}
\newcommand{\eq}[1]{Eq.~(\ref{eq:#1})}


\title{Differentiating information transfer and causal effect}


\author{Joseph T. Lizier}
 \email[]{jlizier@it.usyd.edu.au}
\affiliation{CSIRO Information and Communications Technology Centre,
Locked Bag 17, North Ryde, NSW 1670, Australia}
\affiliation{School of Information Technologies, The University of Sydney,
NSW 2006, Australia}

\author{Mikhail Prokopenko}
\affiliation{CSIRO Information and Communications Technology Centre,
Locked Bag 17, North Ryde, NSW 1670, Australia}



\date{\today}

\begin{abstract}
The concepts of information transfer and causal effect have received much recent attention, yet often the two are not appropriately distinguished and certain measures have been suggested to be suitable for both. We discuss two existing measures, transfer entropy and information flow, which can be used separately to quantify information transfer and causal information flow respectively. We apply these measures to cellular automata on a local scale in space and time, in order to explicitly contrast them and emphasize the differences between information transfer and causality. We also describe the manner in which the measures are complementary, including the circumstances under which the transfer entropy is the best available choice to infer a causal effect. We show that causal information flow is a primary tool to describe the causal structure of a system, while information transfer can then be used to describe the emergent computation in the system.
\end{abstract}


\maketitle


\section{\label{intro}Introduction}

Information transfer is currently a popular topic in complex systems science, with recent investigations spanning cellular automata \cite{liz08a}, biological signaling networks \cite{pahle08,tung07}, and agent-based systems \cite{lung06}.
%
%
In general, information transfer refers to a directional signal or communication of dynamic information from a \textit{source} to a \textit{destination}.
However, the body of literature regarding quantification of information transfer appears to subsume two concepts: \textit{predictive} or \textit{computational information transfer}, and \textit{causal effect} or \textit{information flow}.
That correlation is not causation is well-understood. Yet while authors increasingly consider the notions of information transfer and information flow and how they fit with our understanding of correlation and causality \cite{liang08,lud08,aul08,hlav07,schr00,ay06,lung07b,ish08a}, several questions nag. Is information transfer akin to causal effect? If not, what is the distinction between them? When examining the ``effect" of one variable on another (e.g. between brain regions), should one seek to measure information transfer or causal effect?
Despite the interest in this area, it remains unclear how the notion of information transfer should sit with the concepts of predictive transfer and causal effect.

Predictive transfer refers to the amount of information that a source variable adds to the next state of a destination variable; i.e. ``if I know the state of the source, how much does that help to predict the state of the destination?". This \textit{transferred} information can be thought of as adding to the prediction of an observer, or as being transferred into the computation taking place at the destination \cite{liz08b}; as such, we will also refer to this as the computational perspective.

Causal effect refers to the extent to which the source variable has a direct influence or drive on the next state of a destination variable, i.e. ``if I change the state of the source, to what extent does that alter the state of the destination?". Information from causal effect can be seen to \textit{flow} through the system, like injecting dye into a river \cite{ay06}. In an Aristotelian sense, we restrict our interpretation to \textit{efficient} cause here (e.g. see \cite{vea74}).

Unfortunately, these concepts have become somewhat tangled in discussions of information transfer. Measures for both predictive transfer \cite{schr00} and causal effect \cite{ay06} have been inferred to capture information transfer in general, and measures of predictive transfer have been used to infer causality \cite{sumi08,vej08,lung07b,verd05} with the two sometimes (problematically) directly equated (e.g. \cite{tung07,ish08a,liang08,hlav07,vand07,hung08}).

The notion of information transfer remains cloudy while it is used interchangeably to refer to both concepts.  
Our thesis in this paper is that the concepts of predictive transfer and causal effect are quite distinct: we aim to clarify them and describe the manner in which they should be considered separately.
We argue that the concept of predictive transfer (or the computational perspective) is more closely aligned with the popularly understood notion of information transfer, while causal information flow should be considered separately as a useful notion in its own right.
Using the perspective of information theory (e.g. see \cite{mac03}), we contend that these concepts are properly quantified by the existing measures known as transfer entropy \cite{schr00} and information flow \cite{ay06} respectively, and we use these measures to contrast the concepts.

For this comparison, we examine Cellular Automata (CAs) (e.g. see \cite{wolf02}): discrete dynamical lattice systems involving an array of cells which synchronously update their states as a homogeneous deterministic function of the states of their local neighbors. In particular we focus on Elementary CAs (ECAs), which consist of a one-dimensional array of cells with binary states, with each updated as a function of the previous states of themselves and one neighbor either side (i.e. neighborhood size 3 or range $r=1$). These previous neighborhood states and the recursive chain of their previous neighborhood states form the \textit{past light-cone} of a cell (i.e. the set of all points capable of having a causal effect on it) \cite{sha06}.
CAs provide a well-known example of complex dynamics, since certain rules (e.g. ECA rules 110 and 54 - see \cite{wolf02} regarding the numbering scheme) exhibit emergent structures which are not discernible from their microscopic update functions but which provide the basis for understanding the macroscopic computations carried out in the CAs \cite{mitch98}. These structures include \textit{particles}, which are coherent structures traveling against a background \textit{domain} region. Regular or periodic particles are known as \textit{gliders}. Particles and gliders are important here because they are popularly understood to embody information transfer in the intrinsic computation in the CA \cite{mitch98}. 

In particular, we examine the transfer entropy and information flow measures on a \textit{local} scale in space and time in ECAs, in order to provide an explicit comparison between the two. This is the first presentation and examination of the local information flow.
We demonstrate that transfer entropy as predictive transfer is more closely aligned with the notion of information transfer, since it alone is associated with emergent coherent information transfer structures, i.e. particles in cellular automata. 
We also demonstrate that causality stands separately as a useful concept itself, with information flow identifying causal relations in the domain region of the CA and demonstrating the bounds of influence without being confused by correlations. Additionally, we present parameter settings under which a variant of the transfer entropy may be used to provide an approximation of the information flow.

On the basis of these results, we suggest that information flow should be used first wherever possible in order to establish the set of causal information contributors for a given destination variable. Subsequently, transfer entropy may be used to quantify the information transfer from these causal sources to the destination to study emergent computation in the system.


\section{\label{predictive}Predictive information transfer}

\subsection{\label{transferEntropy}Transfer entropy}
Schreiber presented \textit{transfer entropy} as a measure for information transfer \cite{schr00} in order to address deficiencies in the previous de facto measure, mutual information, the use of which was criticized in this context as a symmetric measure of statically shared information.
Transfer entropy is defined as the deviation from independence (in bits) of the state transition of an information destination \textit{X} from the previous state of an information source \textit{Y}\footnote{The transfer entropy can consider transfer from $l$ previous states of the source $y^{(l)}_{n}$, however here we consider systems where only the previous state of the source is a causal contributor to the destination, so we use $l=1$.}:
\begin{equation}
	T_{Y \rightarrow X} = \sum_{w_n}
    p(w_n)
    \log_{2}{ \frac{ p(x_{n+1}|x^{(k)}_{n},y_{n})}{p(x_{n+1}|x^{(k)}_{n})}}
	\label{eq:te},
\end{equation}
where $n$ is a time index, $x^{(k)}_{n}$ refers to the $k$ states of $X$ up to and including $x_n$, and $w_n$ is the state transition tuple $(x_{n+1},x^{(k)}_{n},y_{n})$. It can be viewed as a \textit{conditional} mutual information, casting it as the average information in the source about the next state of the destination that was not already contained in the destination's past $k$ states.
To ensure that no information in the destination's past is mistaken as transfer here, one should take the limit $k \rightarrow \infty$ though in practice finite-$k$ estimates must be used \cite{liz08a}.
This conditioning on the past makes the transfer entropy a \textit{directional}, \textit{dynamic} measure of information transfer, but it remains a measure of observed (conditional) \textit{correlation} rather than direct effect.
In fact, the transfer entropy is a nonlinear extension of a concept known as the ``Granger causality" \cite{gra69}, the nomenclature for which may have added to the confusion associating information transfer and causal effect.

\subsection{\label{localTransferEntropy}Local transfer entropy}
The transfer entropy is an average (or \textit{expectation value}) of a \textit{local} transfer entropy \cite{liz08a} at each observation $n$, i.e. $T_{Y \rightarrow X} = \left\langle t_{Y \rightarrow X}(n+1) \right\rangle$ where:
\begin{eqnarray}
	t_{Y \rightarrow X}(n+1) = \log_{2}{ \frac{ p(x_{n+1}|x^{(k)}_{n},y_{n})}{p(x_{n+1}|x^{(k)}_{n})}}
	\label{eq:localTE_generic}.
\end{eqnarray}

\begin{figure*}
	\subfigure[Apparent transfer entropy]{\fbox{\label{fig:apparentTE}\includegraphics[width=0.32\textwidth]{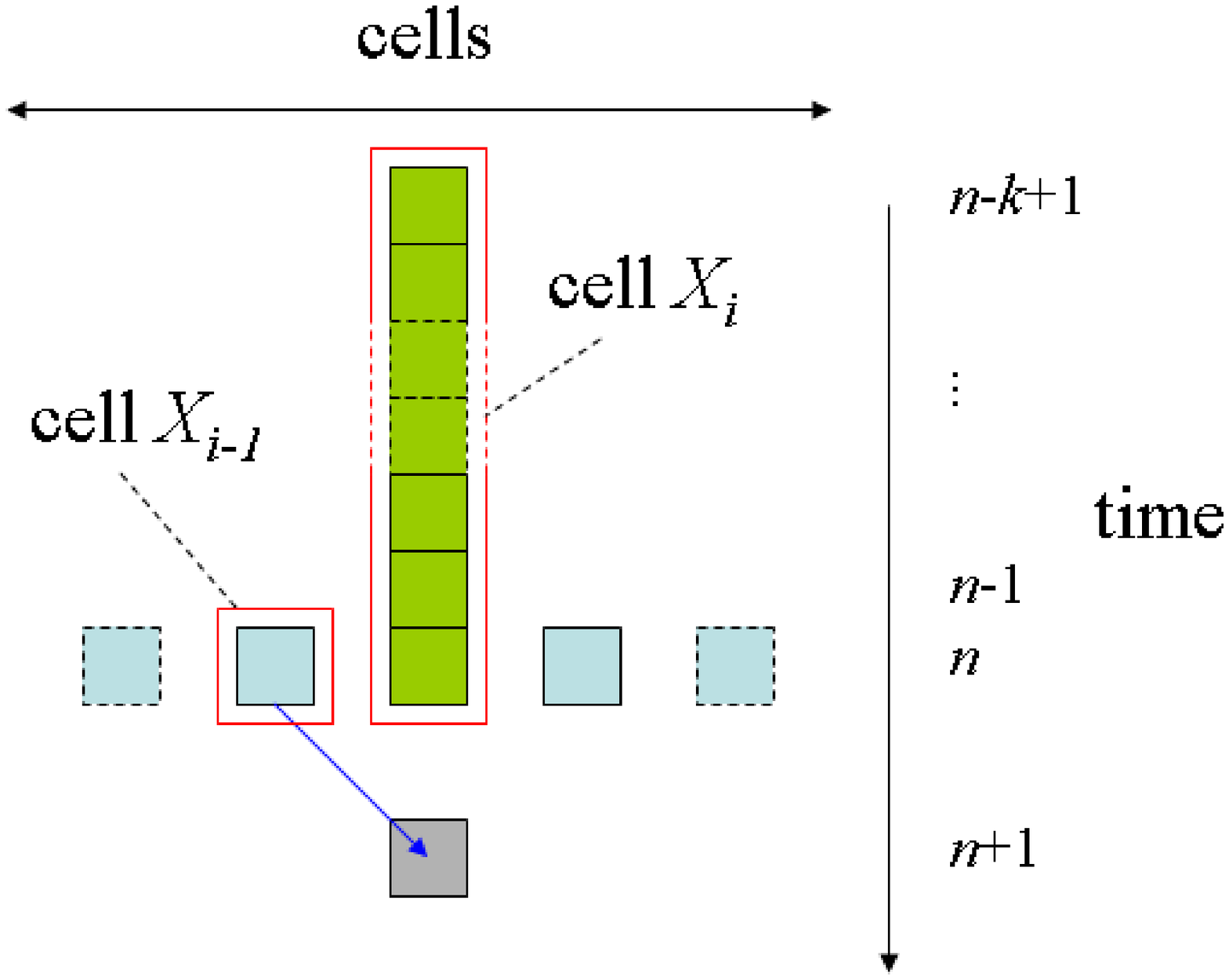}}}
	\subfigure[Complete transfer entropy]{\fbox{\label{fig:completeTE}\includegraphics[width=0.32\textwidth]{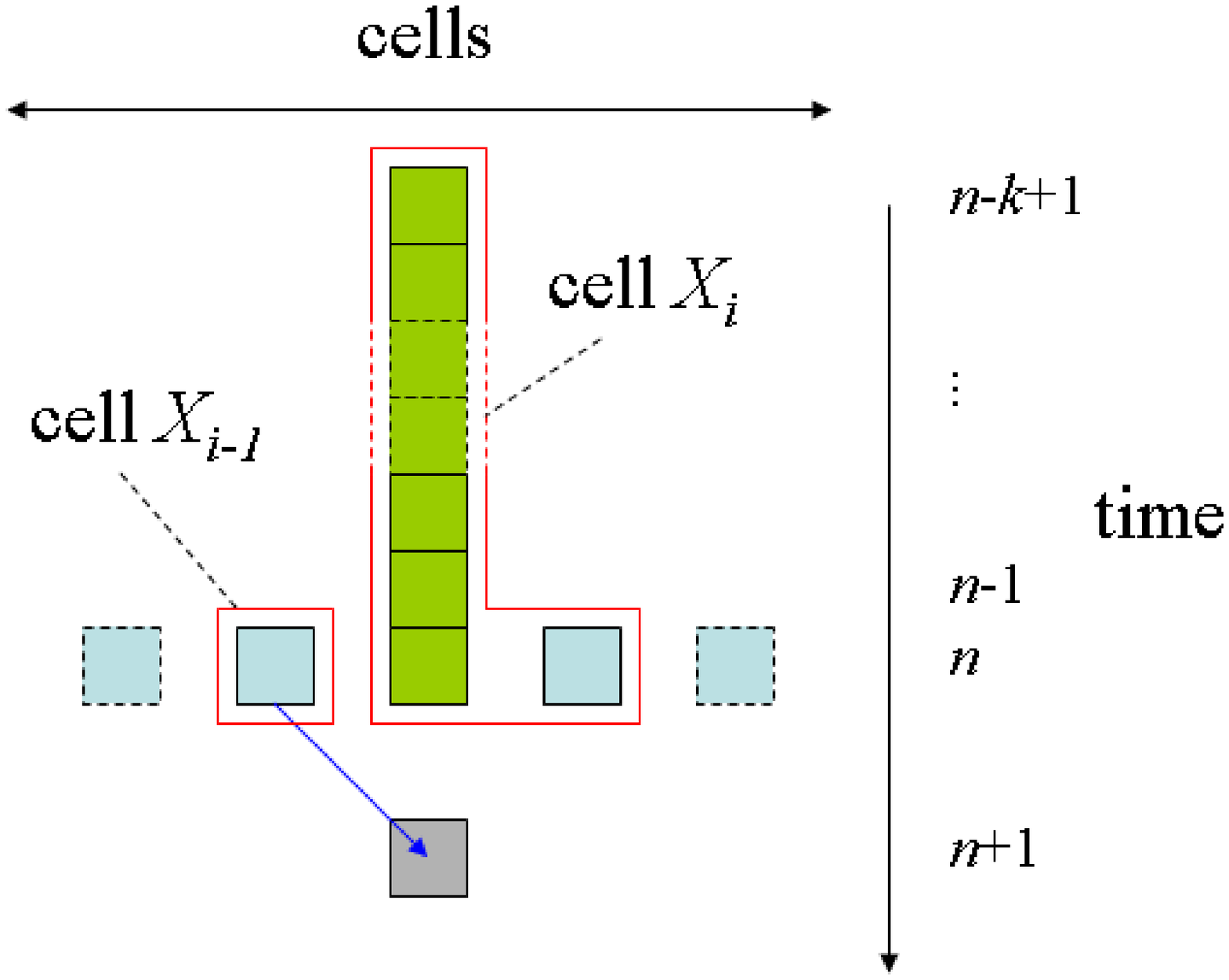}}}
	\subfigure[Information flow]{\fbox{\label{fig:infoFlow}\includegraphics[width=0.32\textwidth]{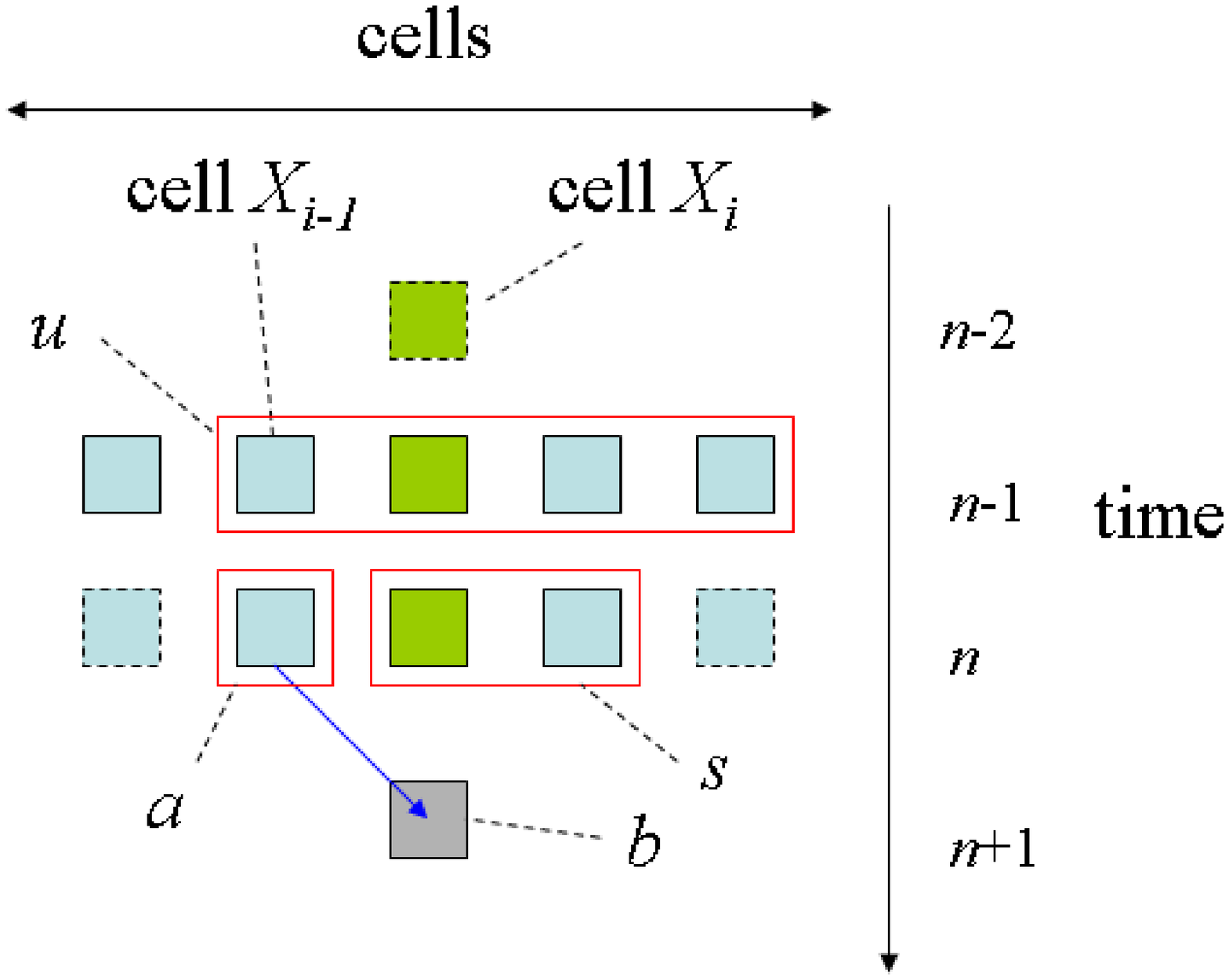}}}
	\caption{\label{fig:diagrams} Measures of information transfer and causality measured across one cell to the right in ECAs. 
			\subref{fig:apparentTE} Apparent transfer entropy: information contained in the source cell $X_{i-1}$ about the next state of the destination cell $X_{i}$ at time time $n+1$ that was not contained in the destination's past.
			\subref{fig:completeTE} Complete transfer entropy: information contained in the source cell $X_{i-1}$ about the next state of the destination cell $X_{i}$ at time time $n+1$ that was not contained in \textit{either} the destination's past \textit{or} the other information contributing cell $X_{i+1}$. As per Section \ref{ateNotCausal}, transfer entropy should only be interpreted as information transfer when measured from within the past light-cone of $x_{i,n}$.
			\subref{fig:infoFlow} Information flow: the contribution of a causal effect from source cell $X_{i-1}$ to the next state of the destination cell $X_{i}$ at time time $n+1$, imposing the previous states of the destination cell and the other information contributing cell $X_{i+1}$; here the source $a=x_{i-1,n}$, the destination $b=x_{i,n+1}$, the imposed contributors are $s=\left\{ x_{i,n}, x_{i+1,n} \right\}$ and the cells blocking a back-door path relative to $(s,a)$ are $u=\left\{ x_{i-1,n-1}, x_{i,n-1}, x_{i+1,n-1}, x_{i+2,n-1} \right\}$.
	}
\end{figure*}

For lattice systems such as CAs with spatially-ordered agents, the local transfer entropy to agent $X_{i}$ from $X_{i-j}$ at time $n+1$ is represented as:
\begin{equation}
	t(i,j,n+1,k) = \log_{2}{ \frac{ p(x_{i,n+1}|x^{(k)}_{i,n},x_{i-j,n})}{p(x_{i,n+1}|x^{(k)}_{i,n})}}
	\label{eq:localTeLattice}.
\end{equation}
The transfer entropy $t(i,j=1,n+1,k)$ to agent $X_{i}$ from $X_{i-1}$ at time $n+1$ is illustrated in \fig{apparentTE}. 
$t(i,j,n,k)$ is defined for every spatiotemporal destination $(i,n)$, for every information channel or direction \textit{j}; sensible values for $j$ correspond to causal information sources, i.e. for CAs, sources within the cell range $|j| \leq r$. We write the average for these lattice systems as $T(j,k) = \left\langle t(i,j,n,k) \right\rangle$.

The transfer entropy may also be conditioned on other possible causal information sources, to eliminate their influence from being attributed to the source in question \textit{Y} \cite{schr00}.
In general, this means conditioning on all sources $Z$ in $X$'s set of causal information contributors $V$ (except for $Y$) with joint state $v_{y,n}$, 
giving the local \textit{complete} transfer entropy \cite{liz08a}:
\begin{eqnarray}
	t^c_{Y \rightarrow X}(n+1,k) = \log_{2}{ \frac{ p(x_{n+1}|x^{(k)}_{n},y_{n},v_{y,n})}{p(x_{n+1}|x^{(k)}_{n},v_{y,n})}}
	\label{eq:localTE_genericComplete}, \\
	v_{y,n} = \left\{ z_n | \forall Z \in V, Z \neq Y,X \right\}
	\label{eq:neighbourhood_generic}.
\end{eqnarray}
For CAs this means conditioning on other sources $v^{r}_{i,j,n}$ within the range $r$ of the destination to obtain \cite{liz08a}:
\begin{eqnarray}
	t^{c}(i,j,n+1,k) = \log_{2}{ \frac
		{p \left( x_{i,n+1}|x^{(k)}_{i,n},x_{i-j,n},v^{r}_{i,j,n} \right)}
		{p \left( x_{i,n+1}|x^{(k)}_{i,n},v^{r}_{i,j,n} \right)}}
	\label{eq:completeTE}, \\
	v^{r}_{i,j,n} = \left\{ x_{i+q,n} | \forall q: -r \leq q \leq +r, q \neq -j, 0 \right\}
	\label{eq:neighbourhood}.
\end{eqnarray}
In deterministic systems (e.g. CAs), complete conditioning renders $t^{c}(i,j,n) \geq 0$ because the source can only add information about the outcome of the destination. Calculations conditioned on no other information contributors (as in \eq{localTeLattice}) are labeled as \textit{apparent} transfer entropy.

Finally, note that the information (or local entropy) $h(i,n+1)$ required to predict the next state of a destination can be decomposed as a sum of \cite{liz08b}:
\begin{itemize}
	\item the information gained from the past of the destination (i.e. the mutual information between the past $x^{(k)}_{i,n}$ and next state $x_{i,n+1}$, known as the active information storage $a(i,n+1,k)$); plus
	\item the information gained from each causal source considered (in arbitrary order) in the context of that past, incrementally conditioning each contribution on the previously considered sources.
\end{itemize}
For example, in ECAs we have:
\begin{eqnarray}
		h(i,n+1) = && a(i,n+1,k) + t(i,j=-1,n+1,k) + \nonumber \\
							 && t^c(i,j=1,n+1,k)
	\label{eq:breakdownEcas},
\end{eqnarray}
In this way, the different forms of the transfer entropy as information transfer can be seen to characterize important components of the total information at the destination.

\section{\label{causal}Causal effect}
It is well-recognized that measurement of causal effect necessitates some type of \textit{perturbation} or \textit{intervention} of the source so as to detect the effect of the intervention on the destination (e.g. see \cite{pearl00,hope05}).
Attempting to infer causality without doing so leaves one measuring correlations of observations, regardless of how directional they may be \cite{ay06}. Here, we adopt the measure information flow for this purpose, and describe how to apply it on a local scale.

\subsection{\label{infoFlow}Information flow}

Following Pearl's probabilistic formulation of causal Bayesian networks \cite{pearl00}, Ay and Polani \cite{ay06} consider how to measure causal information flow via interventional conditional probabilities. For instance, an interventional conditional probability $p(a | \hat{s})$ considers the distribution of $a$ resulting from \textit{imposing} the value of $\hat{s}$. \textit{Imposing} means intervening in the system to set the value of the imposed variable, and is at the essence of the definition of causal information flow.

In a similar fashion to the definition of transfer entropy as the deviation of a destination from \textit{stochastic} independence on the source in the content of the destination's past, Ay and Polani propose the measure \textit{information flow} as the deviation of the destination $B$ from \textit{causal} independence on the source $A$ \textit{imposing} another set of nodes $S$. Mathematically, this is written as:
\begin{eqnarray}
	I_p(A \rightarrow B | \widehat{S}) = \sum_{q} {\hat{p}(q)
		\log_2 { \frac {p(b| \hat{a}, \hat{s})} { \sum_{a'} {p(a' | \hat{s})  p(b| \hat{a}', \hat{s}) } } }
 }
	\label{eq:flow},
\end{eqnarray}
with $q$ representing the tuple $(s,a,b)$ and the modified interventional distribution defined as:
\begin{eqnarray}
	\hat{p}(s,a,b) := p(s) p(a | \hat{s}) p(b| \hat{a}, \hat{s})
	\label{eq:modIntDist}.
\end{eqnarray}

The value of the measure is dependent on the choice of the set of nodes $S$. To obtain the \textit{direct} causal information flow from $A$ to $B$ we must either include all possible other sources in $S$ or at least include enough sources to block all non-immediate directed paths from $A$ to $B$ \cite{ay06}. The minimum to satisfy this is the set of all direct causal sources of $B$ excluding $A$, including any past states of $B$ that are direct causal sources.
For computing direct information flow across one cell to the right in ECAs (see \fig{infoFlow}) where $a = x_{i-1,n}$ and $b = x_{i,n+1}$, this means $S$ includes the immediate past of the destination cell and the previous state of the cell on its right (i.e. $s_{i,n}=\left\{ x_{i,n}, x_{i+1,n} \right\}$). Generalized as $I_p(j)$ for information flow across $j$ cells to the right in any 1D CA, we have:
\begin{eqnarray}
	s_{i,j,n}=\left\{ x_{i,n}, v^r_{i,j,n} \right\}
	\label{eq:sForCAs}.
\end{eqnarray}

Establishing the value of $I_p(A \rightarrow B | \hat{S})$ requires determination of the underlying interventional conditional probabilities. By definition these may be gleaned by observing the results of intervening in the system, however this is not possible in many cases.

One alternative is to use detailed knowledge of the dynamics, in particular the structure of the causal links and possibly the underlying rules of the causal interactions. This also is often not available in many cases, and indeed is often the very goal for which one turned to such analysis in the first place. Regardless, where such knowledge is available it may allow one to make direct inferences, e.g. under complete determination of the observed variable by the imposing set (e.g. $p(b| \hat{a}, \hat{s})$ in ECAs in \fig{infoFlow}), or where the observed variable remains unaffected by the imposition (e.g. $p(a | \hat{s})$ in ECAs in \fig{infoFlow}) allowing one to use the observational probabilities alone independently of the imposed variable.

Furthermore, certain cases exist where one can construct these value from observational probabilities only \cite{ay06}. For example, the ``back-door adjustment" (Section 3.3.1 of \cite{pearl00})\footnote{The back-door adjustment is a sub-case of the ``adjustment for direct causes" \cite{ay06} which is numerically simpler when the set of back-door nodes $U$ is known.} suggests that where a set of nodes $U$ satisfies the ``back-door criteria" relative to $(X,Y)$, i.e. that:
	\begin{enumerate}
		\item ``no node in $U$ is a descendant of" $X$, and
		\item ``$U$ blocks every path between" $X$ and $Y$ that contains a directed causal link into $X$;
	\end{enumerate}
then the interventional conditional probability $p(y|\hat{x})$ is given by:
\begin{eqnarray}
	p(y|\hat{x}) = \sum_{u}{p(y|x,u)p(u)}
	\label{eq:backDoorCondProb}.
\end{eqnarray}
The back-door adjustment could be applied to $p(a | \hat{s})$ in ECAs in \fig{infoFlow} with the set of nodes satisfying the back-door criteria marked there as $u$; for $p(b | \hat{a}, \hat{s})$ the set $u_2 = \left\{ u, x_{i-2,n-1} \right\}$ would be used.
In general, note that the back-door adjustment can only be applied for the information flow in isolation (i.e. without knowledge of the underlying rules of causal interactions) where all relevant combinations are observed (i.e. for $(y,x,u)$ where $p(y,x,u)$ is strictly positive \cite{ay06}).

\subsection{\label{localInfoFlow}Local information flow}
We can define a \textit{local information flow}:
\begin{eqnarray}
	f(a \rightarrow b | \hat{s}) = 
		\log_2 { \frac {p(b| \hat{a}, \hat{s})} { \sum_{a'} {p(a' | \hat{s})  p(b| \hat{a}', \hat{s}) } } }
	\label{eq:localFlow},
\end{eqnarray}
in a similar manner to the localization performed for the transfer entropy. The meaning of the local information flow is slightly different however. Certainly, it is an attribution of local causal effect of $a$ on $b$ were $\hat{s}$ imposed at the given observation $(a,b,s)$. However, one must be aware that $I_p(A \rightarrow B | \hat{S})$ is not the \textit{average} of the local values $f(a \rightarrow b | \hat{s})$. Unlike the transfer entropy, the information flow is averaged over the modified interventional distribution $\hat{p}(s,a,b)$: a product of \textit{interventional} conditional probabilities (see \eq{modIntDist}) which in general does not reduce down to the probability of the given observation $p(s,a,b)$.
For example, it is possible that not all of the tuples $(a,b,s)$ will actually be observed, so averaging over observations would ignore the important contribution that any unobserved tuples provide to the determination of information flow.

For lattice systems such as CAs, we use the notation $f(i,j,n+1)$ to denote the local information flow into agent $X_i$ from the source agent $X_{i-j}$ at time step $n+1$ (i.e. flow across $j$ cells to the right), giving:
\begin{equation}
	f(i,j,n+1) =
	 \log_2 { \frac {p(x_{i,n+1}| \widehat{x_{i-j,n}}, \widehat{s_{i,j,n}})} { d(i,j,n+1) } }
	\label{eq:localFlowLattice},
\end{equation}
\begin{align}
	d(i,j,n+1) = \sum_{x_{i-j,n}'} & p(x_{i-j,n}' | \widehat{s_{i,j,n}}) \nonumber \\
		 & \times \ p(x_{i,n+1}| \widehat{x_{i-j,n}}', \widehat{s_{i,j,n}})
	\label{eq:localFlowLatticeDenominator},
\end{align}
with $s_{i,j,n}$ defined in \ref{eq:sForCAs}.

\section{\label{cas}Application to Cellular Automata}
\begin{figure*}
	\subfigure[Raw CA]{\fbox{\label{fig:54-raw}\includegraphics[width=0.32\textwidth]{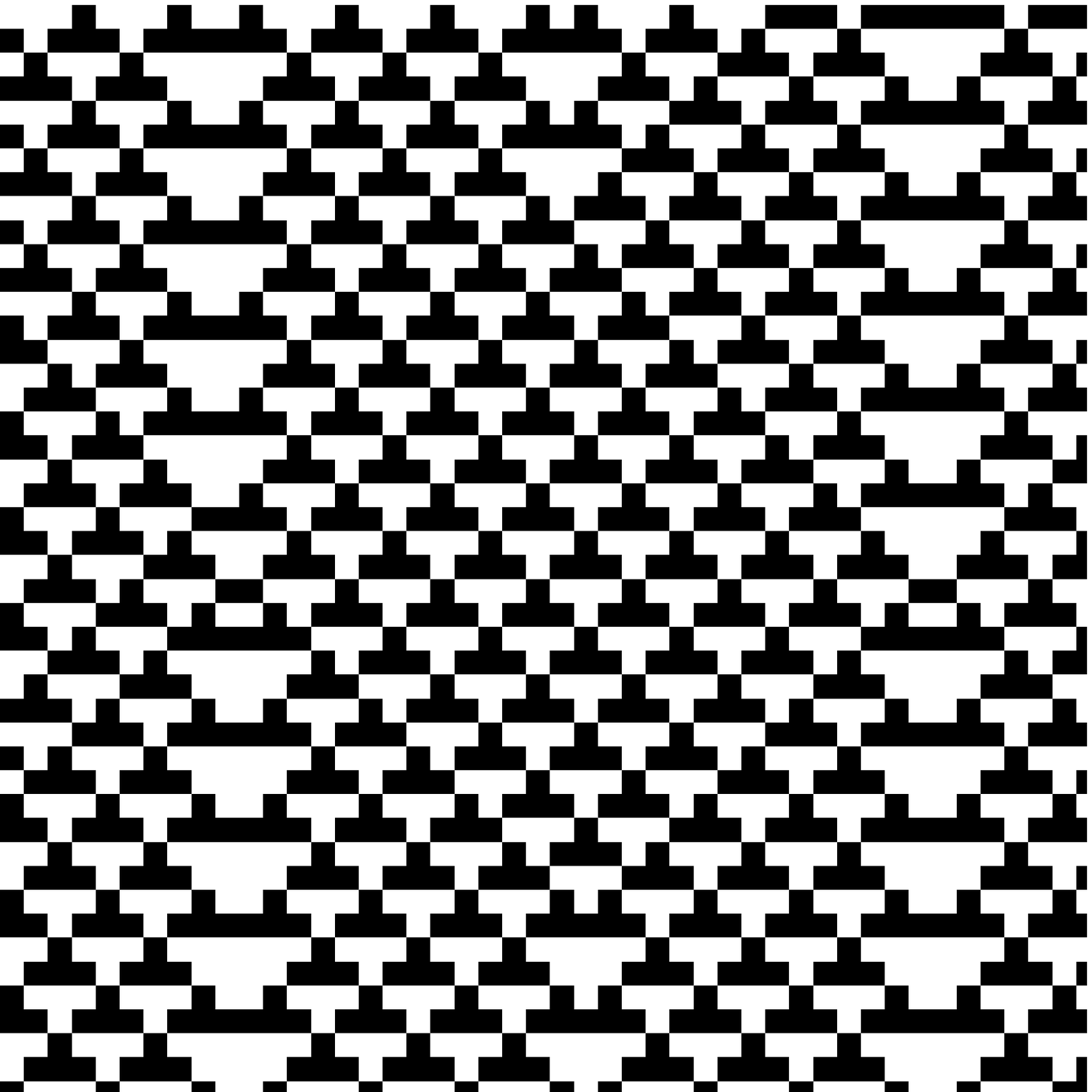}}}
	\subfigure[$f(i,j=1,n)$]{\fbox{\label{fig:54-flow-j1}\includegraphics[width=0.32\textwidth]{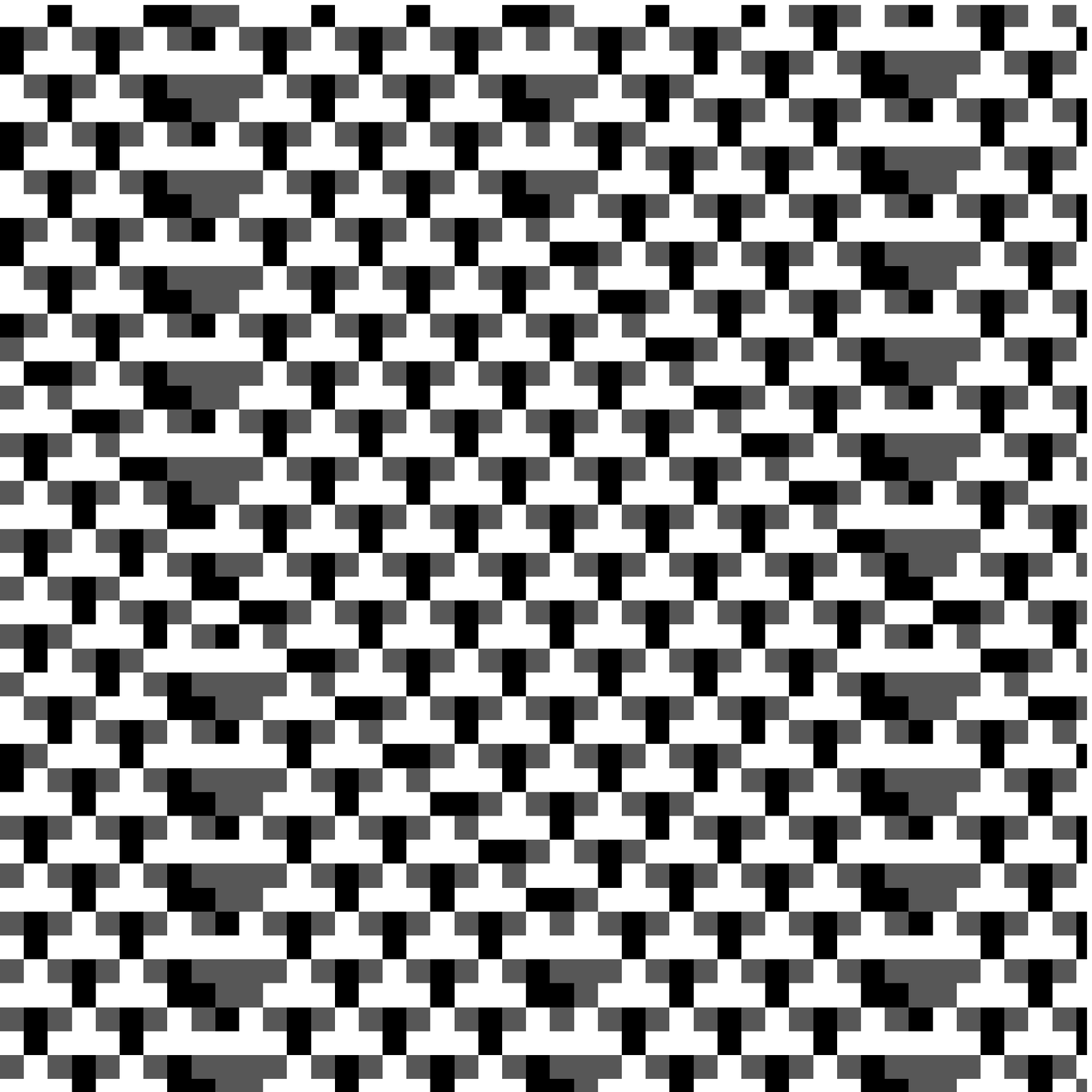}}}
	\subfigure[$t^c(i,j=1,n,k=1)$]{\fbox{\label{fig:54-complete-j1-k1}\includegraphics[width=0.32\textwidth]{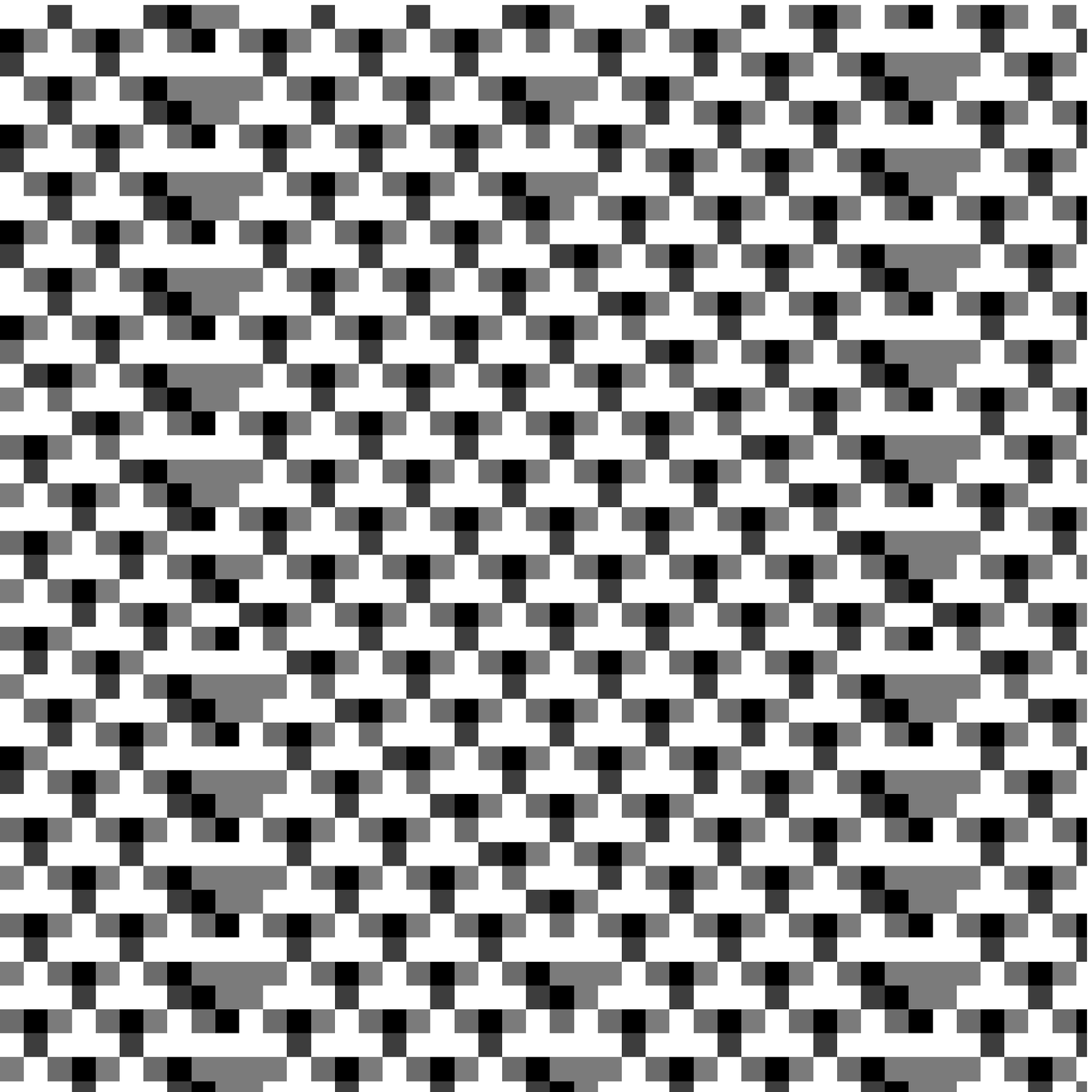}}}
	\subfigure[$t^c(i,j=1,n,k=16)$]{\fbox{\label{fig:54-complete-j1-k16}\includegraphics[width=0.32\textwidth]{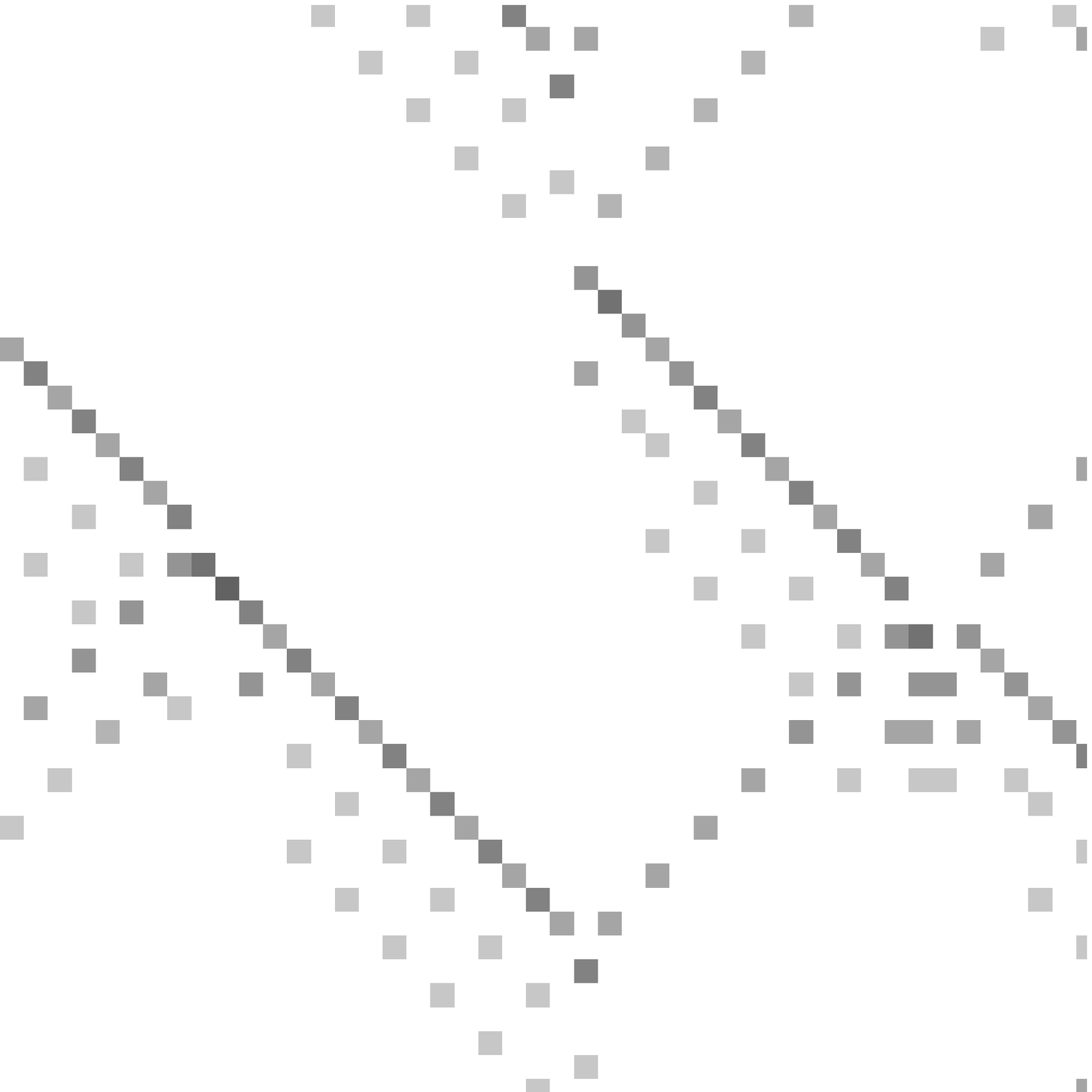}}}
	\subfigure[$t(i,j=1,n,k=16)$]{\fbox{\label{fig:54-apparent-j1-k16-pos}\includegraphics[width=0.32\textwidth]{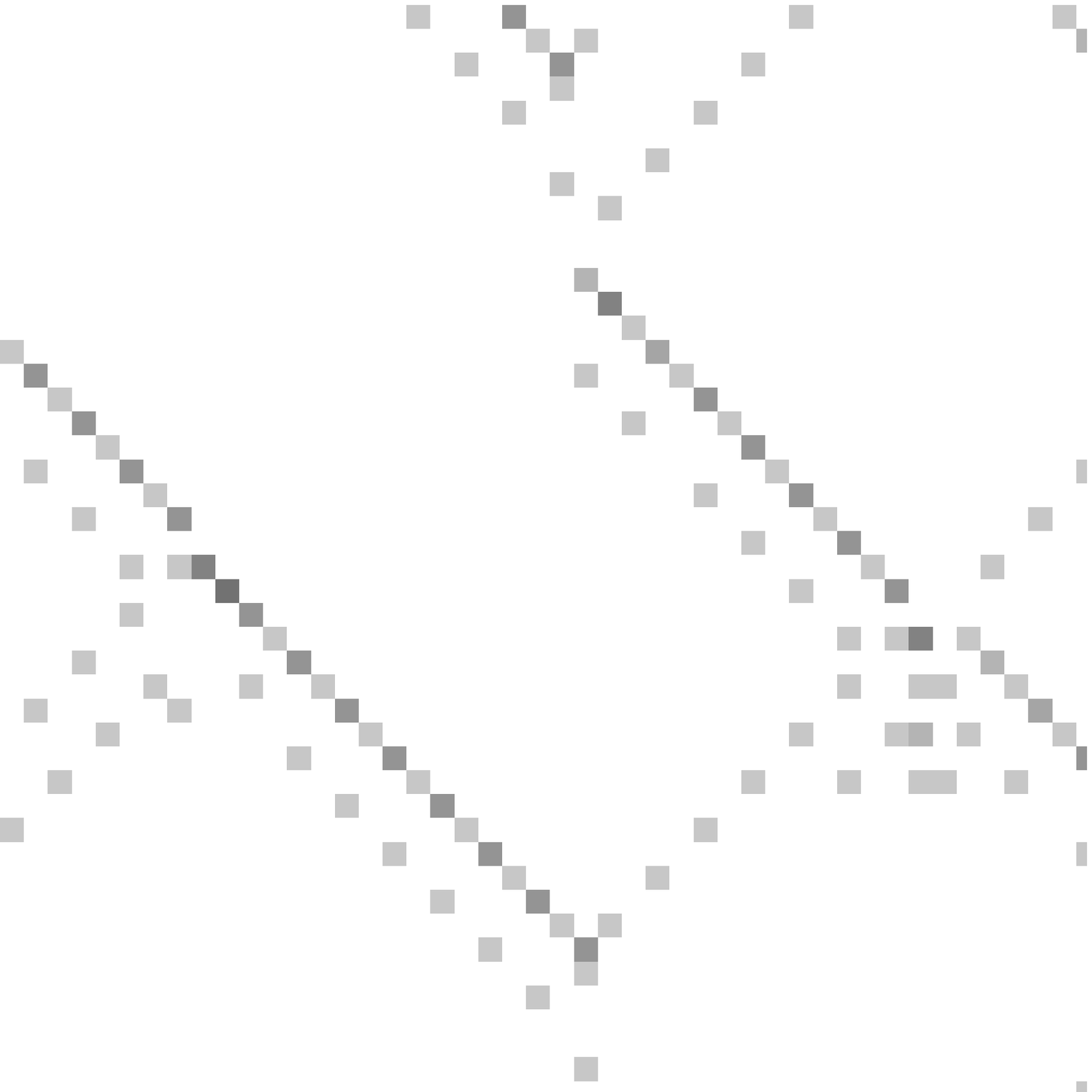}}}
	\subfigure[$t(i,j=2,n,k=16)$]{\fbox{\label{fig:54-apparent-j2-k16-pos}\includegraphics[width=0.32\textwidth]{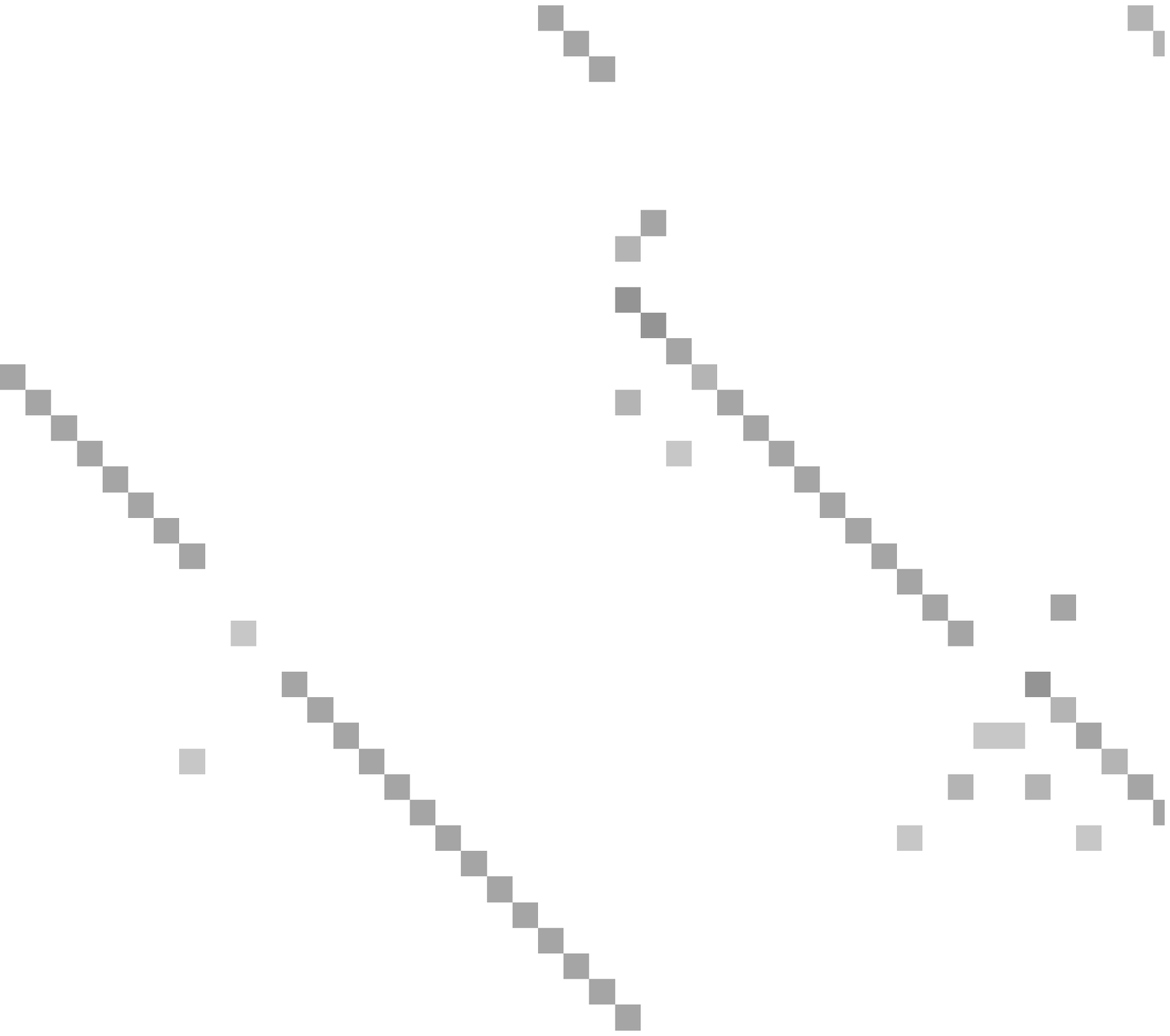}}}

			\caption{\label{fig:54} Local transfer entropy and information flow for raw states of rule 54 in \subref{fig:54-raw} (45 time steps displayed for 45 cells, time increases down the page):
		\subref{fig:54-flow-j1} \textit{Local information flow} across one cell to the right, (all figures gray-scale with 16 levels) with max. 1.07 bits (black);
		\textit{Local complete transfer entropy} across one cell to the right: with past history length $k=1$ in \subref{fig:54-complete-j1-k1} (max. 1.17 bits (black)), and past history length $k=16$ in \subref{fig:54-complete-j1-k16} (max. 9.22 bits (black));
		\textit{Local apparent transfer entropy}, positive values only: across one cell to the right in \subref{fig:54-apparent-j1-k16-pos} (max. 7.93 bits (black)), and across two cells to the right in \subref{fig:54-apparent-j2-k16-pos} (max. 6.00 bits (black)).
	}
\end{figure*}

Here, we apply the local transfer entropy and local information flow to the raw states of ECA rule 54 in \fig{54}. This rule exhibits a (spatially and temporally) periodic background domain, with gliders traveling across the domain and colliding with one another, forming the basis of an emergent intrinsic computation.

Focusing on transfer and flow one step to the right per unit time step, we measure the average transfer values being $T(j=1,k=16) = 0.080$ and $T^c(j=1,k=16) = 0.193$ bits for apparent and complete transfer entropy respectively, and the information flow at $I_p(j = 1) = 0.523$ bits. Much more insight is provided by examining the \textit{local} values of each measure however, and we examine four cases within these results to highlight the differences in the measures and indeed in the concepts of information transfer and causal effect. For measuring the information flow, $p(b | \hat{a}, \hat{s})$ is measured using observations only (unless otherwise stated) to minimize reliance on knowledge of the underlying dynamics.
%

\subsection{\label{domains}Background domains are highly causal}
As an extension of the example of coupled Markov chains in \cite{ay06} to more complex dynamics, we first look at the background domain region of the CA where each cell executes a periodic sequence of states. The four time step period of the (longest) sequences is longer than any one binary-state cell could produce alone - the cells rely on interaction with their neighbors to produce these long sequences. We see that the local transfer entropies $t(i,j=1,n, k=16)$ and $t^c(i,j=1,n, k=16)$ measure vanishing information transfer here in \fig{54-complete-j1-k16} and \fig{54-apparent-j1-k16-pos} \cite{liz08a}, while the local information flow $f(i,j=1,n)$ in \fig{54-flow-j1} measures a periodic pattern of causal effect at similar levels to those in the glider/blinker regions.

\textit{Both results are correct, but from different perspectives}. From a computational perspective, the cells in the domain region are executing information storage processes -- their futures are (almost) completely predictable from their pasts \cite{liz08a}. Note that to achieve these long periods, some of this information is stored in neighbors and retrieved after a few time steps \cite{liz08b} (achieving a stigmergic information storage, similar to \cite{kly04a}). As such, there is vanishing information transfer here. On the other hand, much of the background domain is highly causal because had one imposed values on the sources there the destinations would have changed; hence we find the strong patterns of information flow here. We can also interpret this result by noting that the long periodic sequences in the background domain are underpinned by causal effect between the neighbors.

\subsection{\label{gliders}Gliders distinguished as emergent information transfer}
We then examine the measurements at the gliders, the emergent structures which propagate against the background domain. Here we see that the local transfer entropies $t(i,j=1,n, k=16)$ and $t^c(i,j=1,n, k=16)$ measure strong information transfer in the direction of glider motion in \fig{54-complete-j1-k16} and \fig{54-apparent-j1-k16-pos} \cite{liz08a}, while the local information flow $f(i,j=1,n)$ in \fig{54-flow-j1} measures similar levels of causal effect to those in the background domain.

Again, \textit{both results are correct from different perspectives}. The cell states in the glider region provide strong predictive information about the next states in the direction of glider motion: this is why gliders have long been said to transfer information about the dynamics in one part of the CA to another (as quantified by the local transfer entropy \cite{liz08a}). For this reason, we say that predictive transfer is the concept that more closely aligned with the popularly understood concept of information transfer.
From a causal perspective, the same CA rules executed in the glider are also executed elsewhere in the domain of the CA -- while imposing the source value does indeed have a causal effect on the destination in the gliders, the positive directional information flow here is no greater than levels observed in the domain. The measure certainly captures the causal flow in the gliders, but its localization does not distinguish that from the flow in the domain.

It is possible that a macroscopic formulation of the information flow might distinguish gliders as highly causal macroscopic structures, but certainly (when applied to the same source and destination pair as transfer entropy) as a directional measure of direct local causal effect it does not distinguish these emergent structures.
In this form, the causal perspective focuses on the details or micro-level of the dynamics, whereas the predictive or computational perspective takes a macroscopic view of emergent structures. It is the examination in the context of the past $k$ states that affords this macroscopic view to the transfer entropy. On the other hand, information flow intrinsically cannot consider the context of the past, since imposing on $x_{i-j,n}$ and $s_{i,j,n}$ blocks out the influence of those past $k$ states.

\subsection{\label{ateNotCausal}Information transfer to be measured from causal sources only}
\fig{54-apparent-j2-k16-pos} measures the local apparent transfer entropy $t(i,j=2,n,k=16)$ for two steps to the right per unit time step. This profile is fairly similar to that produced for one step to the right per unit time step. However, this measurement is for \textit{superluminal} transfer, i.e. transfer from outside of the past light-cone of $x_{i,n}$. There should not be a real information transfer here -- what we see in this profile merely reflects a correlation between the purported source and an actual causal source one cell away from the destination. This does not mean that the transfer entropy measure is wrong, merely that it has not been correctly applied here.
It is only causal sources that are present in \eq{breakdownEcas} in contributing the correct information to predict the next state of the destination: \textit{in order to be genuinely interpreted as information transfer, the transfer entropy should only be applied to causal information sources for the given destination.}

To check the correctness of the information flow measure, we apply it here assuming the CA is of neighborhood-5 (i.e. two causal contributors on either side of the destination). As expected, the local information flow profile computes no causal effect across two cells to the right per unit time step (not shown).
Importantly however, note that the information flow could not be measured using observational data alone for either $j=1$ or $j=2$ in neighborhood-5 (since the CA does not produce all of the required $(s,a)$ combinations for computing $p(b|\hat{a},\hat{s})$); specific knowledge about the dynamics was required for the calculation.

Furthermore, measuring the complete transfer entropy $t^c(i,j=2,n,k=16)$ in this neighborhood results in a zero information transfer profile (not shown), since all the required information to predict the next state of the destination is contained within the interior neighborhood for this deterministic system.
This aligns well with the zero result for information flow. Significantly, only the complete transfer entropy is able to make its inference using the available observational data alone, though both measures require the correct neighborhood of other causal contributors to be a subset of those conditioned on or imposed here.

\subsection{\label{completeApproximates}Complete transfer entropy as a next best inference for information flow}
The approximation that the complete transfer entropy provides to the information flow goes beyond similar inference of a lack of influence. Consider the profile of $t^c(i,j=1,n,k=1)$ in \fig{54-complete-j1-k1} -- note how similar it is to the profile of the local information flow in \fig{54-flow-j1}. This is because with the history length $k=1$, the complete transfer entropy measures the information contributed by the source to the destination conditioning out only the information in other causal contributors. The equation for the local complete transfer entropy with $k=1$ (\eq{completeTE}) is indeed very similar to that for the local information flow (\eq{localFlowLattice}), though they are measured over observational and interventional conditional probabilities respectively. Note also that the average value $T^c(j=1,k=1) = 0.521$ bits is almost identical to the information flow $I_p(j=1) = 0.523$ bits.
Where one cannot intervene in the system, and does not have the required observations to use a method such as the back-door adjustment, the local complete transfer entropy could provide next best inference for the local information flow profile. In this case, the history length $k$ should be set to include only the past states of the destination that are causal information contributors to its next state -- \textit{no more, no less}. (For example, in \cite{lung07b} where the elements in Henon maps are causally effected by their previous two states, $k=2$ would be appropriate rather than the use of $k=1$ there). The history length parameter $k$ therefore has an important role in moving the (complete) transfer entropy between measuring information transfer (at large $k$) and approximating causal effect (at minimal $k$).


The complete transfer entropy is therefore a candidate method for \textit{inferring} causal structure in a multi-variate time series in these appropriate conditions, so long as one understands it is neither a direct nor exact \textit{measure} of causal effect. 

Importantly, the complete transfer entropy must condition on (at least) the correct neighborhood of causal sources in order to provide best approximation of the information flow. Since this is exactly what is being searched for in this circumstance, one would in fact need to \textit{build} knowledge of the causal contributors for a given destination by \textit{incrementally} conditioning on previously inferred sources (reminiscent of \eq{breakdownEcas}). This would be done by incrementally selecting the source which provides the most statistically significant transfer entropy conditioned on the previously selected sources (this combines the multi-variate source selection of \cite{tung07} with the complete transfer entropy and the statistical significance tests of \cite{verd05}). Testing this method is left for future work.

Importantly also, while the complete transfer entropy can at least function in the absence of observations spanning all possible combinations of the variables, if crucial combinations are not observed it can give quite incorrect inferences here. For example, consider the classical causal example of a short circuit which causes a fire in the presence of certain conditions (e.g. with inflammable material), while the fire can also be started in other ways (e.g. overturning a lighted oil stove) \cite{mack93}. If one never observes the short circuit in the right conditions, without the other fire triggers, the transfer entropy is in fact unable to infer a causal link from the short circuit to the fire.

\section{\label{discussion}Discussion and Conclusion}

The concepts of information transfer and causal effect have often been confused. In this paper, we have demonstrated the complementary nature of these concepts while emphasizing the distinctions between them.
On an information-theoretical basis, information flow quantifies causal effect using an interventionist perspective, while transfer entropy quantifies information transfer by measuring a (conditional) correlation on a causal channel.
We have explored the subtle yet distinct differences between these concepts using a local scale within cellular automata.

Causal effect is a fundamental micro-level property of a system. Information flow should be used as a primary tool (where possible) to establish the presence of and quantify causal relationships. Where this is not possible (e.g. where one has no ability to intervene in the system, no knowledge of the underlying dynamics, and cannot apply a method such as the back-door adjustment to observational data), then the complete transfer entropy (with history length $k$ set to a minimal value) is an alternate inference technique. The apparent transfer entropy is not applicable here since it cannot discern correlation from causal effect, and neither apparent nor complete transfer entropy with large $k$ is suitable since these measure predictive information transfer rather than direct causal effect.
Note that for both the information flow or complete transfer entropy, it is crucial that they be applied imposing or conditioning the correct set of other causal variables -- the task of building knowledge of this correct set is left for investigation in future work.

Information transfer can then be analyzed in order to gain insight into the emergent computation being carried out by the system. Importantly, the transfer entropy should only be measured for causal information contributors to the destination, otherwise its result cannot be interpreted as information transfer. To do so, both the apparent and complete transfer entropy should be used, with history length $k$ set as large as possible. These are complementary measures which allow one to assess the composition of information storage, transfer and interactions in a system \cite{liz08b}. Information flow is not suitable for the analysis of emergent computation, since in representing causal effect it takes too microscopic a viewpoint, and provides no method for describing the composition of information in the computation.


\begin{acknowledgments}
The authors thank Daniel Polani and Nihat Ay for helpful discussions regarding the nature of the information flow measure and in particular how to estimate it from observational data. JL thanks John Mahoney for discussions regarding measuring transfer entropy from non-causal sources, and the Australian Research Council Complex Open Systems Research Network (COSNet) for a travel grant that partially supported this work.
\end{acknowledgments}

\bibliography{InfoTransferVsCausalEffect}

\begin{thebibliography}{28}
\expandafter\ifx\csname natexlab\endcsname\relax\def\natexlab#1{#1}\fi
\expandafter\ifx\csname bibnamefont\endcsname\relax
  \def\bibnamefont#1{#1}\fi
\expandafter\ifx\csname bibfnamefont\endcsname\relax
  \def\bibfnamefont#1{#1}\fi
\expandafter\ifx\csname citenamefont\endcsname\relax
  \def\citenamefont#1{#1}\fi
\expandafter\ifx\csname url\endcsname\relax
  \def\url#1{\texttt{#1}}\fi
\expandafter\ifx\csname urlprefix\endcsname\relax\def\urlprefix{URL }\fi
\providecommand{\bibinfo}[2]{#2}
\providecommand{\eprint}[2][]{\url{#2}}

\bibitem[{\citenamefont{Lizier et~al.}(2008{\natexlab{a}})\citenamefont{Lizier,
  Prokopenko, and Zomaya}}]{liz08a}
\bibinfo{author}{\bibfnamefont{J.~T.} \bibnamefont{Lizier}},
  \bibinfo{author}{\bibfnamefont{M.}~\bibnamefont{Prokopenko}},
  \bibnamefont{and} \bibinfo{author}{\bibfnamefont{A.~Y.}
  \bibnamefont{Zomaya}}, \bibinfo{journal}{Phys. Rev. E}
  \textbf{\bibinfo{volume}{77}}, \bibinfo{pages}{026110}
  (\bibinfo{year}{2008}{\natexlab{a}}).

\bibitem[{\citenamefont{Pahle et~al.}(2008)\citenamefont{Pahle, Green, Dixon,
  and Kummer}}]{pahle08}
\bibinfo{author}{\bibfnamefont{J.}~\bibnamefont{Pahle}},
  \bibinfo{author}{\bibfnamefont{A.~K.} \bibnamefont{Green}},
  \bibinfo{author}{\bibfnamefont{C.~J.} \bibnamefont{Dixon}}, \bibnamefont{and}
  \bibinfo{author}{\bibfnamefont{U.}~\bibnamefont{Kummer}},
  \bibinfo{journal}{BMC Bioinformatics} \textbf{\bibinfo{volume}{9}},
  \bibinfo{pages}{139} (\bibinfo{year}{2008}).

\bibitem[{\citenamefont{Tung et~al.}(2007)\citenamefont{Tung, Ryu, Lee, and
  Lee}}]{tung07}
\bibinfo{author}{\bibfnamefont{T.~Q.} \bibnamefont{Tung}},
  \bibinfo{author}{\bibfnamefont{T.}~\bibnamefont{Ryu}},
  \bibinfo{author}{\bibfnamefont{K.~H.} \bibnamefont{Lee}}, \bibnamefont{and}
  \bibinfo{author}{\bibfnamefont{D.}~\bibnamefont{Lee}}, in
  \emph{\bibinfo{booktitle}{Proceedings of the Twentieth IEEE International
  Symposium on Computer-Based Medical Systems (CBMS '07), Maribor, Slovenia}},
  edited by \bibinfo{editor}{\bibfnamefont{P.}~\bibnamefont{Kokol}},
  \bibinfo{editor}{\bibfnamefont{V.}~\bibnamefont{Podgorelec}},
  \bibinfo{editor}{\bibfnamefont{D.}~\bibnamefont{Mi\v{c}eti\v{c}-Turk}},
  \bibinfo{editor}{\bibfnamefont{M.}~\bibnamefont{Zorman}}, \bibnamefont{and}
  \bibinfo{editor}{\bibfnamefont{M.}~\bibnamefont{Verli\v{c}}}
  (\bibinfo{publisher}{IEEE}, \bibinfo{address}{Los Alamitos},
  \bibinfo{year}{2007}), pp. \bibinfo{pages}{383--388}.

\bibitem[{\citenamefont{Lungarella and Sporns}(2006)}]{lung06}
\bibinfo{author}{\bibfnamefont{M.}~\bibnamefont{Lungarella}} \bibnamefont{and}
  \bibinfo{author}{\bibfnamefont{O.}~\bibnamefont{Sporns}},
  \bibinfo{journal}{PLoS Comput. Biol.} \textbf{\bibinfo{volume}{2}},
  \bibinfo{pages}{e144} (\bibinfo{year}{2006}).

\bibitem[{\citenamefont{Liang}(2008)}]{liang08}
\bibinfo{author}{\bibfnamefont{X.~S.} \bibnamefont{Liang}},
  \bibinfo{journal}{Phys. Rev. E} \textbf{\bibinfo{volume}{78}},
  \bibinfo{pages}{031113} (\bibinfo{year}{2008}).

\bibitem[{\citenamefont{L\"udtke et~al.}(2008)\citenamefont{L\"udtke, Panzeri,
  Brown, Broomhead, Knowles, Montemurro, and Kell}}]{lud08}
\bibinfo{author}{\bibfnamefont{N.}~\bibnamefont{L\"udtke}},
  \bibinfo{author}{\bibfnamefont{S.}~\bibnamefont{Panzeri}},
  \bibinfo{author}{\bibfnamefont{M.}~\bibnamefont{Brown}},
  \bibinfo{author}{\bibfnamefont{D.~S.} \bibnamefont{Broomhead}},
  \bibinfo{author}{\bibfnamefont{J.}~\bibnamefont{Knowles}},
  \bibinfo{author}{\bibfnamefont{M.~A.} \bibnamefont{Montemurro}},
  \bibnamefont{and} \bibinfo{author}{\bibfnamefont{D.~B.} \bibnamefont{Kell}},
  \bibinfo{journal}{J. R. Soc. Interface} \textbf{\bibinfo{volume}{5}},
  \bibinfo{pages}{223} (\bibinfo{year}{2008}).

\bibitem[{\citenamefont{Auletta et~al.}(2008)\citenamefont{Auletta, Ellis, and
  Jaeger}}]{aul08}
\bibinfo{author}{\bibfnamefont{G.}~\bibnamefont{Auletta}},
  \bibinfo{author}{\bibfnamefont{G.~F.~R.} \bibnamefont{Ellis}},
  \bibnamefont{and} \bibinfo{author}{\bibfnamefont{L.}~\bibnamefont{Jaeger}},
  \bibinfo{journal}{J. R. Soc. Interface} \textbf{\bibinfo{volume}{5}},
  \bibinfo{pages}{1159} (\bibinfo{year}{2008}).

\bibitem[{\citenamefont{Hlav\'{a}\v{c}kov\'{a}-Schindler
  et~al.}(2007)\citenamefont{Hlav\'{a}\v{c}kov\'{a}-Schindler, Palu\v{s},
  Vejmelka, and Bhattacharya}}]{hlav07}
\bibinfo{author}{\bibfnamefont{K.}~\bibnamefont{Hlav\'{a}\v{c}kov\'{a}-Schindl%
er}}, \bibinfo{author}{\bibfnamefont{M.}~\bibnamefont{Palu\v{s}}},
  \bibinfo{author}{\bibfnamefont{M.}~\bibnamefont{Vejmelka}}, \bibnamefont{and}
  \bibinfo{author}{\bibfnamefont{J.}~\bibnamefont{Bhattacharya}},
  \bibinfo{journal}{Physics Reports} \textbf{\bibinfo{volume}{441}},
  \bibinfo{pages}{1} (\bibinfo{year}{2007}).

\bibitem[{\citenamefont{Schreiber}(2000)}]{schr00}
\bibinfo{author}{\bibfnamefont{T.}~\bibnamefont{Schreiber}},
  \bibinfo{journal}{Phys. Rev. Lett.} \textbf{\bibinfo{volume}{85}},
  \bibinfo{pages}{461} (\bibinfo{year}{2000}).

\bibitem[{\citenamefont{Ay and Polani}(2008)}]{ay06}
\bibinfo{author}{\bibfnamefont{N.}~\bibnamefont{Ay}} \bibnamefont{and}
  \bibinfo{author}{\bibfnamefont{D.}~\bibnamefont{Polani}},
  \bibinfo{journal}{Adv. Complex Syst.} \textbf{\bibinfo{volume}{11}},
  \bibinfo{pages}{17} (\bibinfo{year}{2008}).

\bibitem[{\citenamefont{Lungarella et~al.}(2007)\citenamefont{Lungarella,
  Ishiguro, Kuniyoshi, and Otsu}}]{lung07b}
\bibinfo{author}{\bibfnamefont{M.}~\bibnamefont{Lungarella}},
  \bibinfo{author}{\bibfnamefont{K.}~\bibnamefont{Ishiguro}},
  \bibinfo{author}{\bibfnamefont{Y.}~\bibnamefont{Kuniyoshi}},
  \bibnamefont{and} \bibinfo{author}{\bibfnamefont{N.}~\bibnamefont{Otsu}},
  \bibinfo{journal}{Int. J. Bifurcation Chaos} \textbf{\bibinfo{volume}{17}},
  \bibinfo{pages}{903} (\bibinfo{year}{2007}).

\bibitem[{\citenamefont{Ishiguro et~al.}(2008)\citenamefont{Ishiguro, Otsu,
  Lungarella, and Kuniyoshi}}]{ish08a}
\bibinfo{author}{\bibfnamefont{K.}~\bibnamefont{Ishiguro}},
  \bibinfo{author}{\bibfnamefont{N.}~\bibnamefont{Otsu}},
  \bibinfo{author}{\bibfnamefont{M.}~\bibnamefont{Lungarella}},
  \bibnamefont{and}
  \bibinfo{author}{\bibfnamefont{Y.}~\bibnamefont{Kuniyoshi}},
  \bibinfo{journal}{Phys. Rev. E} \textbf{\bibinfo{volume}{77}},
  \bibinfo{pages}{026216} (\bibinfo{year}{2008}).

\bibitem[{\citenamefont{Lizier et~al.}(2008{\natexlab{b}})\citenamefont{Lizier,
  Prokopenko, and Zomaya}}]{liz08b}
\bibinfo{author}{\bibfnamefont{J.~T.} \bibnamefont{Lizier}},
  \bibinfo{author}{\bibfnamefont{M.}~\bibnamefont{Prokopenko}},
  \bibnamefont{and} \bibinfo{author}{\bibfnamefont{A.~Y.}
  \bibnamefont{Zomaya}}, \emph{\bibinfo{title}{A framework for the local
  information dynamics of distributed computation in complex systems}}
  (\bibinfo{year}{2008}{\natexlab{b}}), \bibinfo{note}{arXiv:0811.2690},
  \urlprefix\url{http://arxiv.org/abs/0811.2690}.

\bibitem[{\citenamefont{Veatch}(1974)}]{vea74}
\bibinfo{author}{\bibfnamefont{H.~B.} \bibnamefont{Veatch}},
  \emph{\bibinfo{title}{Aristotle: A contemporary appreciation}}
  (\bibinfo{publisher}{Indiana University Press},
  \bibinfo{address}{Bloomington}, \bibinfo{year}{1974}).

\bibitem[{\citenamefont{Sumioka et~al.}(2007)\citenamefont{Sumioka, Yoshikawa,
  and Asada}}]{sumi08}
\bibinfo{author}{\bibfnamefont{H.}~\bibnamefont{Sumioka}},
  \bibinfo{author}{\bibfnamefont{Y.}~\bibnamefont{Yoshikawa}},
  \bibnamefont{and} \bibinfo{author}{\bibfnamefont{M.}~\bibnamefont{Asada}}, in
  \emph{\bibinfo{booktitle}{Proceedings of the 6th IEEE International
  Conference on Development and Learning (ICDL 2007), London}}
  (\bibinfo{publisher}{IEEE}, \bibinfo{year}{2007}), pp.
  \bibinfo{pages}{264--269}.

\bibitem[{\citenamefont{Vejmelka and Palus}(2008)}]{vej08}
\bibinfo{author}{\bibfnamefont{M.}~\bibnamefont{Vejmelka}} \bibnamefont{and}
  \bibinfo{author}{\bibfnamefont{M.}~\bibnamefont{Palus}},
  \bibinfo{journal}{Phys. Rev. E} \textbf{\bibinfo{volume}{77}},
  \bibinfo{pages}{026214} (\bibinfo{year}{2008}).

\bibitem[{\citenamefont{Verdes}(2005)}]{verd05}
\bibinfo{author}{\bibfnamefont{P.~F.} \bibnamefont{Verdes}},
  \bibinfo{journal}{Phys. Rev. E} \textbf{\bibinfo{volume}{72}},
  \bibinfo{pages}{026222} (\bibinfo{year}{2005}).

\bibitem[{\citenamefont{Van~Dijck et~al.}(2007)\citenamefont{Van~Dijck,
  Van~Vaerenbergh, and Van~Hulle}}]{vand07}
\bibinfo{author}{\bibfnamefont{G.}~\bibnamefont{Van~Dijck}},
  \bibinfo{author}{\bibfnamefont{J.}~\bibnamefont{Van~Vaerenbergh}},
  \bibnamefont{and} \bibinfo{author}{\bibfnamefont{M.~M.}
  \bibnamefont{Van~Hulle}}, in \emph{\bibinfo{booktitle}{Proceedings of the
  International Conference on Artificial Neural Networks (ICANN 2007), Porto,
  Portugal}}, edited by \bibinfo{editor}{\bibfnamefont{J.~M.~d.}
  \bibnamefont{S{\'a}}}, \bibinfo{editor}{\bibfnamefont{L.~A.}
  \bibnamefont{Alexandre}},
  \bibinfo{editor}{\bibfnamefont{W.}~\bibnamefont{Duch}}, \bibnamefont{and}
  \bibinfo{editor}{\bibfnamefont{D.}~\bibnamefont{Mandic}}
  (\bibinfo{publisher}{Springer-Verlag}, \bibinfo{address}{Berlin/Heidelberg},
  \bibinfo{year}{2007}), vol. \bibinfo{volume}{4669} of
  \emph{\bibinfo{series}{Lecture Notes in Computer Science}}, pp.
  \bibinfo{pages}{159--168}.

\bibitem[{\citenamefont{Hung and Hu}(2008)}]{hung08}
\bibinfo{author}{\bibfnamefont{Y.-C.} \bibnamefont{Hung}} \bibnamefont{and}
  \bibinfo{author}{\bibfnamefont{C.-K.} \bibnamefont{Hu}},
  \bibinfo{journal}{Phys. Rev. Lett.} \textbf{\bibinfo{volume}{101}},
  \bibinfo{pages}{244102} (\bibinfo{year}{2008}).

\bibitem[{\citenamefont{MacKay}(2003)}]{mac03}
\bibinfo{author}{\bibfnamefont{D.~J.} \bibnamefont{MacKay}},
  \emph{\bibinfo{title}{Information Theory, Inference, and Learning
  Algorithms}} (\bibinfo{publisher}{Cambridge University Press},
  \bibinfo{address}{Cambridge}, \bibinfo{year}{2003}).

\bibitem[{\citenamefont{Wolfram}(2002)}]{wolf02}
\bibinfo{author}{\bibfnamefont{S.}~\bibnamefont{Wolfram}},
  \emph{\bibinfo{title}{A New Kind of Science}} (\bibinfo{publisher}{Wolfram
  Media}, \bibinfo{address}{Champaign, IL, USA}, \bibinfo{year}{2002}).

\bibitem[{\citenamefont{Shalizi et~al.}(2006)\citenamefont{Shalizi, Haslinger,
  Rouquier, Klinkner, and Moore}}]{sha06}
\bibinfo{author}{\bibfnamefont{C.~R.} \bibnamefont{Shalizi}},
  \bibinfo{author}{\bibfnamefont{R.}~\bibnamefont{Haslinger}},
  \bibinfo{author}{\bibfnamefont{J.-B.} \bibnamefont{Rouquier}},
  \bibinfo{author}{\bibfnamefont{K.~L.} \bibnamefont{Klinkner}},
  \bibnamefont{and} \bibinfo{author}{\bibfnamefont{C.}~\bibnamefont{Moore}},
  \bibinfo{journal}{Phys. Rev. E} \textbf{\bibinfo{volume}{73}},
  \bibinfo{pages}{036104} (\bibinfo{year}{2006}).

\bibitem[{\citenamefont{Mitchell}(1998)}]{mitch98}
\bibinfo{author}{\bibfnamefont{M.}~\bibnamefont{Mitchell}}, in
  \emph{\bibinfo{booktitle}{Non-Standard Computation}}, edited by
  \bibinfo{editor}{\bibfnamefont{T.}~\bibnamefont{Gramss}},
  \bibinfo{editor}{\bibfnamefont{S.}~\bibnamefont{Bornholdt}},
  \bibinfo{editor}{\bibfnamefont{M.}~\bibnamefont{Gross}},
  \bibinfo{editor}{\bibfnamefont{M.}~\bibnamefont{Mitchell}}, \bibnamefont{and}
  \bibinfo{editor}{\bibfnamefont{T.}~\bibnamefont{Pellizzari}}
  (\bibinfo{publisher}{VCH Verlagsgesellschaft}, \bibinfo{address}{Weinheim},
  \bibinfo{year}{1998}), pp. \bibinfo{pages}{95--140}.

\bibitem[{\citenamefont{Granger}(1969)}]{gra69}
\bibinfo{author}{\bibfnamefont{C.~W.~J.} \bibnamefont{Granger}},
  \bibinfo{journal}{Econometrica} \textbf{\bibinfo{volume}{37}},
  \bibinfo{pages}{424} (\bibinfo{year}{1969}).

\bibitem[{\citenamefont{Pearl}(2000)}]{pearl00}
\bibinfo{author}{\bibfnamefont{J.}~\bibnamefont{Pearl}},
  \emph{\bibinfo{title}{Causality: Models, Reasoning, and Inference}}
  (\bibinfo{publisher}{Cambridge University Press},
  \bibinfo{address}{Cambridge}, \bibinfo{year}{2000}).

\bibitem[{\citenamefont{Hope and Korb}(2005)}]{hope05}
\bibinfo{author}{\bibfnamefont{L.~R.} \bibnamefont{Hope}} \bibnamefont{and}
  \bibinfo{author}{\bibfnamefont{K.~B.} \bibnamefont{Korb}},
  \bibinfo{type}{Tech. Rep.} \bibinfo{number}{2005/176},
  \bibinfo{institution}{Clayton School of Information Technology, Monash
  University} (\bibinfo{year}{2005}).

\bibitem[{\citenamefont{Klyubin et~al.}(2004)\citenamefont{Klyubin, Polani, and
  Nehaniv}}]{kly04a}
\bibinfo{author}{\bibfnamefont{A.~S.} \bibnamefont{Klyubin}},
  \bibinfo{author}{\bibfnamefont{D.}~\bibnamefont{Polani}}, \bibnamefont{and}
  \bibinfo{author}{\bibfnamefont{C.~L.} \bibnamefont{Nehaniv}}, in
  \emph{\bibinfo{booktitle}{Proceedings of the Ninth International Conference
  on the Simulation and Synthesis of Living Systems (ALife IX), Boston, USA}},
  edited by \bibinfo{editor}{\bibfnamefont{J.}~\bibnamefont{Pollack}},
  \bibinfo{editor}{\bibfnamefont{M.}~\bibnamefont{Bedau}},
  \bibinfo{editor}{\bibfnamefont{P.}~\bibnamefont{Husbands}},
  \bibinfo{editor}{\bibfnamefont{T.}~\bibnamefont{Ikegami}}, \bibnamefont{and}
  \bibinfo{editor}{\bibfnamefont{R.~A.} \bibnamefont{Watson}}
  (\bibinfo{publisher}{MIT Press}, \bibinfo{address}{Cambridge, MA, USA},
  \bibinfo{year}{2004}), pp. \bibinfo{pages}{563--568}.

\bibitem[{\citenamefont{Mackie}(1993)}]{mack93}
\bibinfo{author}{\bibfnamefont{J.~L.} \bibnamefont{Mackie}}, in
  \emph{\bibinfo{booktitle}{Causation}}, edited by
  \bibinfo{editor}{\bibfnamefont{E.}~\bibnamefont{Sosa}} \bibnamefont{and}
  \bibinfo{editor}{\bibfnamefont{M.}~\bibnamefont{Tooley}}
  (\bibinfo{publisher}{Oxford University Press}, \bibinfo{address}{New York,
  USA}, \bibinfo{year}{1993}).

\end{thebibliography}

\end{document}